\documentclass[prl,twocolumn,showpacs,amsmath,amssymb,superscriptaddress,nofootinbib]{revtex4-1}

\usepackage[utf8]{inputenc}
\usepackage{epsfig}
\usepackage{graphicx}
\usepackage{dcolumn}
\usepackage{bm}
\usepackage{color}
\usepackage{url}
\usepackage{ulem}
\usepackage{multirow}
\usepackage{enumitem}
\usepackage{CJK}
\usepackage[bookmarksnumbered,bookmarksopen,colorlinks,citecolor=red,linkcolor=blue]{hyperref}

\newcommand{\pT} {p_{\rm T}}

\newcommand {\phia}	{\phi_{\alpha}}
\newcommand {\phib}	{\phi_{\beta}}

\newcommand {\vres}	{v_{2,{\rm res}}}

\newcommand {\dg}	{\Delta\gamma}

\newcommand {\phires}	{\phi_{\rm res}}

\usepackage{tikz,xcolor,hyperref}
\definecolor{lime}{HTML}{A6CE39}
\DeclareRobustCommand{\orcidicon}{
	\begin{tikzpicture}
	\draw[lime, fill=lime] (0,0) 
	circle [radius=0.16] 
	node[white] {{\fontfamily{qag}\selectfont \tiny ID}};
	\draw[white, fill=white] (-0.0625,0.095) 
	circle [radius=0.007];
	\end{tikzpicture}
	\hspace{-2mm}
}

\foreach \x in {A, ..., Z}{%
	\expandafter\xdef\csname orcid\x\endcsname{\noexpand\href{https://orcid.org/\csname orcidauthor\x\endcsname}{\noexpand\orcidicon}}
}

\newcommand {\mean}[1]	{\langle #1\rangle}

\begin{document}
\begin{CJK}{UTF8}{gbsn}
\title{Isolation of photon-nuclear interaction backgrounds in the search for the chiral magnetic effect in relativistic heavy-ion collisions}
\author{Jing Gu (顾婧)}
\author{Jinhui Chen (陈金辉)}
\author{Jie Zhao (赵杰)}
	\address{Key Laboratory of Nuclear Physics and Ion-beam Application (MOE), Institute of Modern Physics, Fudan University, Shanghai 200433, China}
	\address{Shanghai Research Center for Theoretical Nuclear Physics, NSFC and Fudan University, Shanghai 200438, China}

\date{\today}

\begin{abstract}
The chiral magnetic effect (CME) in relativistic heavy-ion collisions originates from a chirality imbalance among quarks within metastable QCD vacuum domains and may be linked to $CP$ violation, which is believed to play a crucial role in the matter-antimatter asymmetry of the universe. Over the past two decades, extensive experimental efforts at RHIC and the LHC have been devoted to the search for evidence of the CME. Recent advances have greatly improved our understanding of background contributions that can mimic CME-like signals. In particular, analyses utilizing techniques designed to suppress flow-related backgrounds indicate that the CME signal at RHIC, if present, is small. To further investigate potential background sources, particularly those associated with strong electromagnetic fields, we estimate the contribution from coherent photon-nuclear interactions. These interactions are driven by intense electromagnetic fields produced in ultrarelativistic heavy-ion collisions, with cross sections that scale with the field strength. Notably, the polarization of the incident photons is aligned with the electric field, which is oriented along the impact parameter direction and perpendicular to the magnetic field. Consequently, such processes can generate charge-dependent correlations that mimic key features of the CME signal, yet originate from different physics mechanisms and are distinct from flow-induced backgrounds. In this study, we quantitatively assess the influence of these coherent photon-nuclear interactions on the precision measurement of the CME, aiming to improve the separation of the genuine CME signal from these background contributions.

\end{abstract}
\maketitle


\section{Introduction}  
Metastable domains of fluctuating topological charges can alter quark chirality, leading to local parity $(P)$ violation and charge-parity $(CP)$ violation in quantum chromodynamics (QCD)~\cite{Lee:1974ma,Kharzeev:1998kz,Kharzeev:1999cz}.
This would lead to electric charge separation in the presence of a strong magnetic field, a phenomenon known as the chiral magnetic effect (CME)~\cite{Kharzeev:1998kz,Kharzeev:1999cz,Kharzeev:2007jp,Fukushima:2008xe}. 
Such a magnetic field, reaching up to $10^{18}$ G, may arise in non-central relativistic heavy-ion collisions due to fast moving spectator protons at early times~\cite{Kharzeev:2007jp,Fukushima:2008xe,Asakawa:2010bu,Bzdak:2011yy,ALICE:2019sgg,STAR:2019clv,STAR:2023jdd,Shen:2025unr}.
Although a finite CME signal is generally expected~\cite{Kharzeev:1999cz,Kharzeev:2007jp}, precise quantitative predictions remain elusive~\cite{Muller:2010jd}, going beyond mere order-of-magnitude estimates, despite extensive theoretical efforts over the last decade (see recent reviews~\cite{Kharzeev:2013ffa,Kharzeev:2015znc,Huang:2015oca,KharzeevLiaoNPR}).
On the experimental side, extensive efforts have been devoted to searching for charge separation induced by CME at the Relativistic Heavy-Ion Collider (RHIC)~\cite{STAR:2009wot,STAR:2009tro,STAR:2014uiw,STAR:2020gky,STAR:2021pwb,STAR:2023gzg,STAR:2023ioo,STAR:2025uxv,STAR:2025vhs} and the Large Hadron Collider (LHC)~\cite{ALICE:2012nhw,CMS:2016wfo,ALICE:2017sss,ALICE:2020siw} (also see reviews~\cite{Kharzeev:2024zzm,Zhao:2018ixy,Zhao:2018skm,Zhao:2019hta,Li:2020dwr,Chen:2024aom,Feng:2025yte}), 
including a dedicated isobar collision program at RHIC~\cite{Voloshin:2010ut,Skokov:2016yrj,STAR:2021mii}.

The primary observable for measuring the charge separation induced by the CME is the three-point correlator~\cite{Voloshin:2004vk}, defined as
$\gamma \equiv \langle \cos(\phi_\alpha + \phi_\beta - 2\psi) \rangle,$
where $\phi_\alpha$ and $\phi_\beta$ are the azimuthal angles of the particles $\alpha$ and $\beta$, respectively, and $\psi$ denotes the azimuthal angle of the spectator (SP) or the participant plane (PP), determined by the beam direction and the average transverse positions of the spectator or participant nucleons.
To reduce charge-independent background contributions (e.g., those arising from global momentum conservation), the difference between opposite-sign (OS) and same-sign (SS) correlations is often used:
$\Delta\gamma \equiv \gamma_{\mathrm{OS}} - \gamma_{\mathrm{SS}},$
where “OS” (“SS”) denotes particle pairs with electric charges of opposite-sign (same-sign).
A possible CME signal can be quantified by the first-order sine modulation coefficient, $a_1$, in the final-state azimuthal distributions of positive ($+$) and negative ($-$) charged hadrons: $\frac{dN_\pm}{d\phi_\pm} \propto 1 \pm 2a_1 \sin(\phi_\pm - \psi) + 2v_2 \cos 2(\phi_\pm - \psi) + \cdots,$
which leads to a contribution of $\Delta\gamma = 2a_1^2$~\cite{Voloshin:2004vk}.

However, several background sources can also contribute to the observed $\Delta\gamma$~\cite{PhysRevC.107.L031902}.
The most significant among them is the flow-induced background, which arises from the coupling between elliptic flow and the decay of resonances or clusters, as described in~\cite{Voloshin:2004vk,Wang:2009kd,Wang:2016iov}:
\begin{equation}
	\begin{split}
		\dg_{\rm bkgd}^{\rm flow}&= \mean{\cos(\phia+\phib-2\psi)}, \\
	&= \mean{\cos(\phia+\phib-2\phires)}\vres\,,
	\label{eqbkg}
	\end{split}
\end{equation}
where $\phia$ and $\phib$ are the azimuthal angle of the decay daughters.
$\phires$ is the azimuthal angle of the resonance/cluster, and $\vres$ is its elliptic flow coefficient.
The elliptic flow anisotropy is generated from strong interactions that convert the initial geometric anisotropy of the overlapping participant zone into momentum-space anisotropy of final-state hadrons~\cite{Ollitrault:1992bk}.
The geometry of the participant region is correlated with that of the spectators, which in turn generates the electromagnetic field.
Therefore, flow-induced backgrounds are inherently correlated with the CME signal, making them particularly difficult to disentangle.
In recent years, most experimental efforts have focused on suppressing flow-induced backgrounds~\cite{STAR:2023gzg,STAR:2023ioo,STAR:2025uxv,STAR:2025vhs}.
New results from RHIC Au+Au collisions indicate that, after subtracting flow backgrounds, a possible CME signal remains at the level of $\approx$5-10\%, with a significance of 2-3$\sigma$ compared to inclusive measurements~\cite{STAR:2021pwb,STAR:2020gky}.

To further understand possible backgrounds, a natural question arises:
Is there any background that is directly correlated with the electromagnetic field and mimics the CME signal?
Coherent photon-nuclear interactions that produce pairs of particles represent such a source, as they are strongly aligned with the electromagnetic field.
In this study, we estimate the impact of this background on precision measurements of the CME signal.
Specifically, we first calculate the coherent $\rho^0$ cross section in semi-central heavy-ion collisions,
then estimate its contribution to $\Delta\gamma_{\rm bkgd}$, and compare it with the inclusive $\Delta\gamma$ measurements.

\section{Coherent $\rho^{0}$ production}  
In relativistic heavy-ion collisions, the fast-moving nuclei are accompanied by intense photon fluxes because of their large electric charge and Lorentz-contracted electromagnetic fields.
These fields are strong enough to induce both photon-photon and photon-nuclear interactions~\cite{Bertulani:2005ru,Klein:2016yzr,Klein:2020fmr}.
In photon-nuclear interactions, virtual photons emitted by one nucleus can fluctuate into a quark-antiquark ($q\bar{q}$) pair, scatter off the other nucleus (via two gluon exchange, the low order Pomeron exchange) and emerge as vector mesons (e.g., $\rho$, $\omega$, $\phi$, $J/\psi$, etc.).
These processes are typically studied in ultra-peripheral collisions (UPCs), where the impact parameter b is larger than twice the nuclear radius ($R_{A}$) and no hadronic interactions occur.
Numerous interesting results on particle productions such as $e^{+}+e^{-}$, $\pi^{+}+\pi^{-}$ and vector mesons $\rho, \omega, J/\psi$ have been reported in UPCs~\cite{Klein:1999qj,STAR:2002caw,ALICE:2012yye,CMS:2016itn,ATLAS:2017fur,STAR:2019wlg,STAR:2022wfe,Zhao:2022dac,Brandenburg:2025one}.

The photon-nuclear interaction can also occur in hadronic heavy-ion collisions (where $b<2R_{A}$). 
Significant excess yields of $J/\psi$ and dielectrons at very low transverse momentum ($p_{T}<$ 100 MeV/$c$) have been observed by the ALICE Collaboration~\cite{ALICE:2015mzu} and the STAR Collaborations~\cite{STAR:2018ldd}.
These excesses cannot be fully explained by conventional hadronic production mechanisms.
Interestingly, they exhibit features consistent with coherent photon-nuclear interactions.
Theoretical calculations of coherent photon-nuclear processes in hadronic heavy-ion collisions show good agreement with experimental measurements~\cite{Klusek-Gawenda:2015hja,GayDucati:2018who,Zha:2017jch},
indicating that photon-nuclear contributions in such collisions are non-negligible.
Here, ``coherent" refers to interactions in which the photon couples to the entire nucleus as a whole, 
while ``incoherent" interactions involve coupling to individual nucleons.
In this study, we focus exclusively on the coherent $\rho^{0}$ component. The other processes, such as $\omega, \phi, J/\psi$ etc, are negligible compared to $\rho^{0}$.
The incoherent contribution is expected to be less than 50\% of the coherent one. Experimental data show that it has a negligible impact on azimuthal correlations~\cite{STAR:2022wfe,STAR:2019wlg}.

The detailed calculation of the total cross section for coherent photon-nuclear production of a vector meson has been developed in~\cite{Klein:1999qj,Klein:2016yzr}, 
which can be expressed as an integral over the photon energy spectrum~\cite{Klein:1999qj,Klein:2016yzr}: 
\begin{eqnarray}
	\begin{split}
	\sigma(AA\rightarrow AAV)=2\int dk \frac{dN_{\gamma}(k)}{dk}\sigma(\gamma A \rightarrow VA), \\
	\end{split}
\label{eq:AAcs}
\end{eqnarray}
where the photon flux $\frac{dN_{\gamma}(k)}{dk}$ can be calculated using the Weizsäcker-Williams approach,  
and $\sigma(\gamma A \rightarrow VA)$ is the photon-nuclear interaction cross section. This approach has been widely used in the calculation of coherent vector meson production in UPC.  
To extend the calculation to hadronic heavy-ion collisions, the photon flux needs to be modified by taking into account the collision geometry~\cite{Klusek-Gawenda:2015hja,GayDucati:2018who,Zha:2017jch}. 

The electromagnetic field of a relativistic nucleus can be approximated as a flux of quasi-real virtual photons.
The photon flux can be described by the general expression~\cite{GayDucati:2018who}:
\begin{eqnarray}
	\begin{split}
	\frac{d^{3}N(\omega,b)}{d\omega d^{2}b} = \frac{Z^{2}\alpha}{\pi^{2}\omega}\left| \int_{0}^{\infty} dk_{\perp}k^{2}_{\perp} \frac{F(k)}{k^{2}}J_{1}(bk_{\perp})\right|^{2},   \\
		k^{2} = \frac{\omega}{\gamma}^{2} +k_{\perp}^{2},  \gamma = \sqrt{s_{NN}}/(2m_{p}), 
	\end{split}
\label{eq:bkgd}
\end{eqnarray}
where $\alpha \approx$ 1/137 is the fine structure constant, Z is the nuclear charge, $\omega$ is the photon energy, k and $k_{\perp}$ are the photon momentum and transverse momentum. $\gamma$ is the Lorentz factor. 
The form factor $F(k^{2})$ is the Fourier transform of the nucleus charge distribution. 
In the calculation of the photon flux in the UPC, where the impact parameter is large, 
the nucleus is often treated as a point-like charge distribution, with the form factor taken as $F(k^{2})=1$. 
Under this approximation, the photon flux can be given by a simple formula
\begin{equation}
\frac{dN^{3}(\omega,b)}{d\omega d^{2}b} 
=  \frac{Z^{2}\alpha x^{2}}{\pi^{2}kb^{2}}K^{2}_{1}(x),
\end{equation}
with $x=\omega b/\gamma$. In hadronic heavy-ion collisions, a more realistic nuclear form factor must be considered.
Although the Woods-Saxon distribution accurately describes the nuclear charge density, it does not yield an analytic expression for the form factor.
A commonly used approximation involves a hard-sphere nuclear density convoluted with a Yukawa potential of range a=0.7 fm.
In this case, the form factor can be expressed as~\cite{Klein:1999qj} 
\begin{eqnarray}
    \begin{split}
		F(k) = \frac{4\pi\rho_{0}}{Ak^{3}} (\sin(kR_{A}) - kR_{A}\cos(kR_{A}))\times\frac{1}{1+a^{2}k^{2}},
    \end{split}
\label{eq:bkgd}
\end{eqnarray}
where A is the mass number, $\rho_{0}$ is the nuclear density.

\begin{figure}[hbt]
	\centering
	\includegraphics[width=0.49\textwidth]{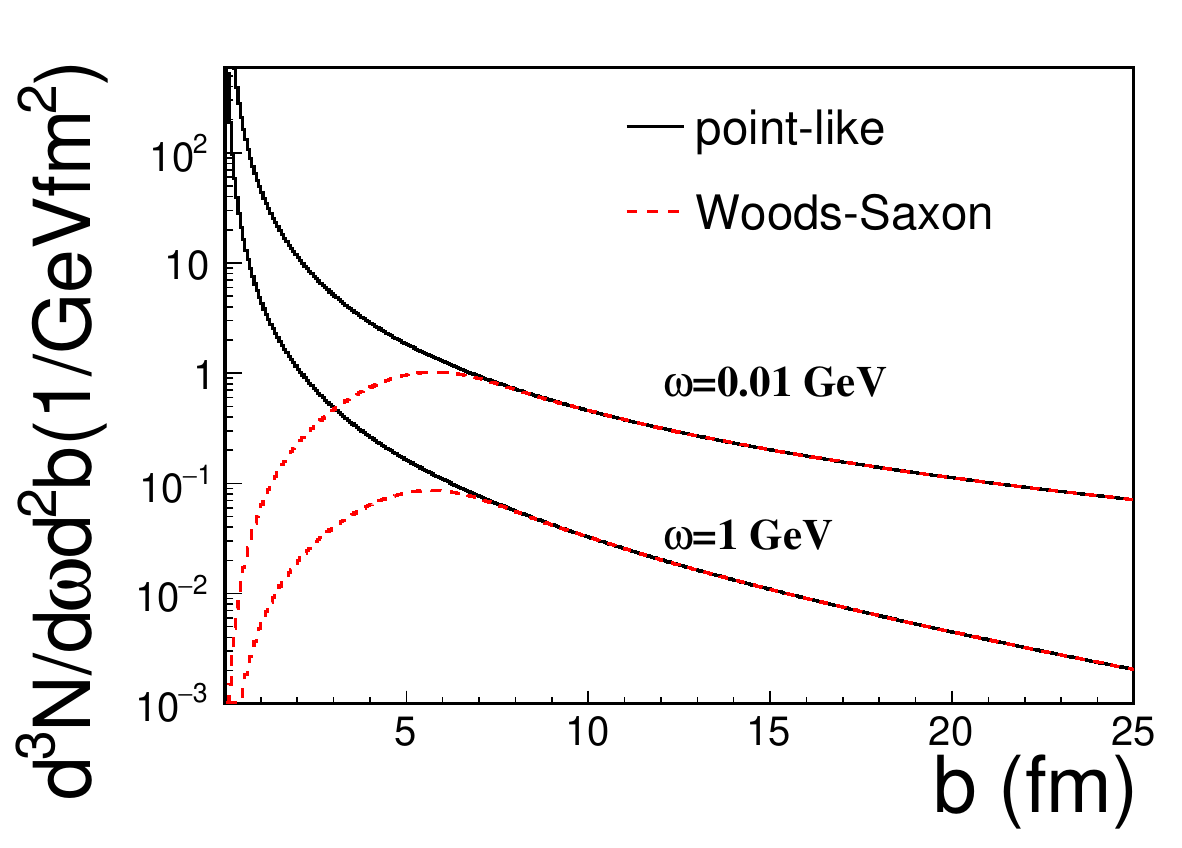}
\caption[]{(Color online) Comparison of the photon flux as a function of impact parameter b with the point-like and realistic form factor at photon energy of $\omega=0.01$ and 1 GeV in Au+Au collisions.
}
\label{CMEphotonEng}
\end{figure}

Figure.~\ref{CMEphotonEng} shows a comparison of the photon flux computed using point-like and realistic Woods-Saxon form factors. 
At large impact parameter photons flux with point-like and realistic Woods-Saxon form factor is consistent with each other. 
In the small b region, which is the relevant range of the hadronic heavy-ion collisions, the point-like charge distribution is no longer a reasonable description. 
Therefore, in the present calculation the realistic Woods-Saxon form factor is used.

To account for the collision geometry in hadronic heavy-ion collisions, the effective photon flux is calculated following~\cite{Klusek-Gawenda:2015hja,GayDucati:2018who} as
\begin{eqnarray}
    \begin{split}
		N^{eff}(\omega,b) = \int d^{2}b_{1} N(\omega,b_{1})\frac{\theta(R_{A}-b_{2})\times\theta(b_{1}-R_{A})}{\pi R^{2}_{A}}, 
    \end{split}
\label{eq:effFlux}
\end{eqnarray}
where $b_1$ and $b_2$ are the transverse distances from the production point to the centers of the photon emitter and target nuclei, respectively.
The $\theta(R_{A}-b_{2})$ ensures that only the photons reach the nuclei target and contribute. 
Similarly, the $\theta(b_{1}-R_{A})$ condition is to ensure that the photons reach the nuclei overlap region was not included. 
There are different considerations of the collision geometry, for example, weather or not consider the overlap region. 
Based on~\cite{Klusek-Gawenda:2015hja}, the calculation with Eq.\ref{eq:effFlux} can well reproduce the experimental data. 
Therefore, this method is used in our estimation.

The next step is to evaluate the coherent photon-nuclear cross section ($\sigma(\gamma A \rightarrow VA)$), the second part of Eq.\ref{eq:AAcs}. 
The coherent photon-nuclear cross section was calculated by~\cite{Klein:1999qj}:
\begin{eqnarray}
    \begin{split}
        \sigma(\gamma A \rightarrow VA) = \left. \frac{d\sigma(\gamma A \rightarrow VA)}{dt} \right|_{t=0} \int dt\,|F(t)|^{2}, \\
		\left. \frac{d\sigma(\gamma A \rightarrow VA)}{dt} \right|_{t=0} = \frac{\alpha\sigma^{2}_{tot}(VA)}{4f_{\pi}}, 
    \end{split}
\label{eq:gAcs}
\end{eqnarray}
where $\left. \frac{d\sigma}{dt} \right|_{t=0}$ is the differential cross section at zero momentum transfer and $F(t)$ is the nuclear form factor. 
This expression is derived under the vector meson dominance (VMD) model, optical theorem, and the assumption of coherent scattering.
The integral over $|F(t)|^2$ accounts for the suppression of large momentum transfers due to the finite nuclear size. 
Here, the nuclear form factor $F(t)$ uses the same function derived from the charge distribution to obtain a good approximation. 
The total vector meson-nuclear scattering cross section can be obtained via the Glauber model as~\cite{Klein:1999qj}:
\begin{eqnarray}
    \begin{split}
		\sigma_{tot}(VA) = \int d^{2}\vec{r}(1-e^{-\sigma_{tot}(Vp)T_{AA}(\vec{r})}),  \\
		\sigma^{2}_{tot}(Vp) =\left. 16\pi\frac{d\sigma(Vp\rightarrow Vp)}{dt} \right|_{t=0}, \\ 
		\left. \frac{d\sigma(Vp\rightarrow Vp)}{dt} \right|_{t=0} = \left. \frac{f^{2}_{v}}{4\pi\alpha} \frac{d\sigma(\gamma p\rightarrow V p)}{dt} \right|_{t=0},
    \end{split}
\label{eq:VAcs}
\end{eqnarray}
where the nuclear thickness function $T_{\rm AA}$ can be calculated using the classical Glauber model. 
Quantum-mechanical Glauber calculations, which resum multiple scatterings and incorporate interference and nuclear shadowing effects, 
typically yield cross sections about 15\% larger than those from classical Glauber models.
Following~\cite{Klusek-Gawenda:2015hja}, we adopt the classical Glauber formalism in this study.
$f_{v}$ is the vector meson-photon coupling constant.
The $\gamma p \rightarrow Vp$ cross section can be extracted from the experimental data, we used the parametrization from~\cite{Klein:1999qj,Klein:2016yzr}. 

\begin{figure}[hbt]
	\centering
	\includegraphics[width=0.49\textwidth]{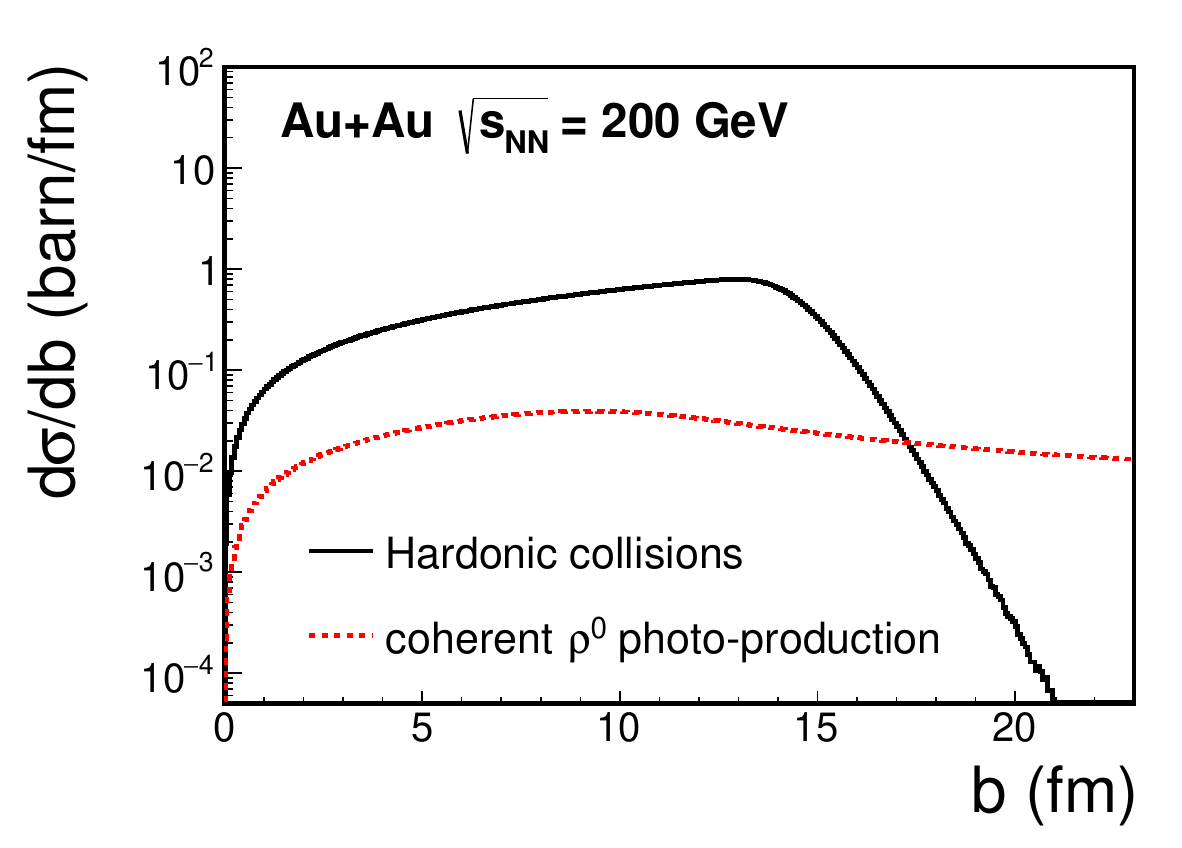}
\caption[]{(Color online) The dashed line show the total coherent $\rho^{0}$ photon-production cross section as function of the impact parameter b in the hadronic heavy-ion collisions. For comparison, the total hadronic cross section from the MC Glauber calculation is shown in solid line. 
}
\label{CME-gA}
\end{figure}

Figure~\ref{CME-gA} shows the estimated total coherent $\rho^{0}$ photon-production cross section as a function of the impact parameter b in the hadronic heavy-ion collisions.
In semi-central Au+Au collisions at $\sqrt{s_{\rm NN}}$= 200 GeV, the coherent cross section $\sigma(\rm Au+Au\rightarrow Au+Au+\rho^{0})$ is about $~10\%$ of the total hadronic cross section from the Monte-Carlo (MC) Glauber calculation. 
The rapidity distribution of the produced vector mesons is determined by the photon energy via the relation $\rm y=ln\frac{2k}{m_{V}}$, where k is the photon energy, $m_{V}$ is the vector meson mass. 
The extracted yield of coherent $\rho^{0}$ in the mid-rapidity ($|y|<1$) is $N_{\rm coherent~\rho^{0}}=0.03$ for 20-50$\%$ centrality Au+Au collisions at $\sqrt{s_{\rm NN}}$= 200 GeV, corresponding to approximately $\approx0.1\%$ of the total hadronic $\rho^{0}$ yield~\cite{Shen:2024eeb}. 
This contribution is typically negligible in hadronic heavy-ion collisions. 
However, given that the CME signal is also expected to be a very small effect~\cite{STAR:2021pwb}, we evaluated the potential impact of coherent $\rho^{0}$ production as a background for CME-related observables.

The photon polarization is aligned with the electric field vector, which points along the impact parameter direction and is perpendicular to the magnetic field~\cite{STAR:2021pwb,STAR:2020gky,Li:2019yzy,Wu:2022exl}.
Under the assumption of helicity conservation (no-helicity flip)~\cite{Schilling:1969um}, the vector meson produced retains the linear polarization of the incident photon.
The decay of the vector meson into two spinless daughter particles is described by the angular distribution~\cite{Schilling:1969um,Wu:2022exl}:
\begin{eqnarray}
    \begin{split}
		\frac{d^{2}N}{d\rm \cos\theta d\phi} = \frac{3}{8\pi}\rm \sin^{2}\theta(1+\cos(2\phi)), 
    \end{split}
\label{eq:VAdecay}
\end{eqnarray}
where the decay angles $\theta$ and $\phi$ are the polar and azimuthal angles, respectively.
This distribution implies a preferential emission of the decay products along the direction of polarization,
hence to the impact parameter direction and perpendicular to the magnetic field direction.
Taking into account the finite nuclear size and realistic density distribution, Ref.~\cite{Wu:2022exl} estimated that for coherent $\rho\rightarrow\pi^{+}+\pi^{-}$
decays, the average correlation between the decay pion with the impact parameter direction is $\langle\cos(2\phi-2\psi)\rangle$=0.38 in the $20-50\%$ centrality of Au+Au collisions at $\sqrt{s_{\rm NN}}$ = 200 GeV.
This preferential orientation of decay angles induces charge-dependent background correlations in the $\dg$ observable that resemble the CME signal but are distinct from flow-induced effects. Specifically, 
\begin{equation}
	\begin{split}
		\dg_{\rm bkgd}^{\rm coherent \rho^{0}}& = \mean{\cos(\phia+\phib-2\psi)},  \\
		&=\mean{\cos(\phia-\phib)\times\cos(2\phib-2\psi)}\,
	\end{split}
	\label{eqgA}
\end{equation}
where $\phia, \phib$ are the azimuthal angles of the decay daughters. 
In the rest frame of the $\rho^{0}$ meson, the two decay pions are emitted back-to-back.
For coherent $\rho^{0}$ meson production, the transverse momentum is limited to $\pT$ $\le \rm \hbar/R_{A} \approx 100$ MeV, and $\pi^{\pm}$ $\cos(\phia-\phib)$ is still $\approx-1$. 
As mentioned above, $\cos(2\phib-2\psi)=0.38$ was used in this estimation. 
Therefore, the estimated contribution from coherent $\rho^{0}$ photon production is $\dg= -0.38$, which is roughly an order of magnitude larger (in absolute value) than the contribution from hadronic $\rho^{0}$ decays~\cite{Wang:2016iov}. 
This negative value implies that the experimentally measured $\dg$ is smaller than its true value when coherent $\rho^{0}$ production is taken into account.

The total contribution of coherent $\rho^{0}$ production to inclusive $\dg$ can be estimated as
\begin{equation}
	\begin{split}
		\mean{\dg_{\rm bkgd}^{\rm coherent \rho^{0}}} = & \rm \mean{\frac{N^{coherent \rho^{0}}}{N^{+}\times N^{-}}}\times\mean{\cos(\phia+\phib-2\psi)} \\
														& \times \mean{\cos2(\psi-\psi_{2})} , 
	\end{split}
	\label{eqgA}
\end{equation}
where $\rm \langle \frac{N^{\mathrm{coherent}\ \rho^{0}}}{N^{+}N^{-}} \rangle$ denotes the multiplicity dilution factor.
$\psi_{2}$ is the experimentally measured second-order event-plane angle.
The factor $\langle \cos 2(\psi-\psi_{2}) \rangle$ accounts for event-by-event fluctuations that affect the azimuthal orientation of the electromagnetic fields, which is approximately $0.65$ for $20-50\%$ centrality Au+Au collisions at $\sqrt{s_{NN}}=200$ GeV according to Ref.~\cite{Bloczynski:2012en}.
From MC Glauber simulation, the total multiplicity of charge particle is $N_{ch}\approx150$ in the pseudo-rapidity range of $|\eta|<0.5$ for the 20-50$\%$ centrality of Au+Au collisions at $\sqrt{s_{\rm NN}}$= 200 GeV.
Here we use $N^{+}=N^{-}=150$ in the pseudo-rapidity range of $|\eta|<1$~\cite{Wang:2016iov}.
The estimated contribution from coherent $\rho^{0}$ photon production is
\begin{equation}
	\mean{\dg_{\rm bkgd}^{\rm coherent \rho^{0}}}=\frac{0.03}{150 \times 150}\times (-0.38) \times 0.65 \approx -0.33 \times 10^{-6}.
    \label{eqgA}
\end{equation}
Compared with the inclusive measurement of $\dg=1.89\times10^{-4}$ and the potential CME fraction of $f_{\rm CME}=14.7\pm5\%$~\cite{STAR:2021pwb}, the coherent contribution of $\rho^{0}$ is around $-0.2\%$ compared to the inclusive $\dg$. 
Therefore, current experimental measurements of the possible CME fraction may be underestimated by about $0.2\%$.

\section{Discussion and Summary}  
Recent experimental analyses at RHIC have significantly reduced flow-induced backgrounds in the search for the chiral magnetic effect (CME), revealing a small but finite residual signal. 
To improve the precision of such measurements, it is essential to evaluate other potential backgrounds, especially those correlated with strong electromagnetic fields. In this study, we estimate the contribution to the three-point correlator $\Delta\gamma$ from coherent $\rho^0$ production via photon-nuclear interactions. 
These processes, characterized by strong alignment with the electric field, lead to charge-dependent azimuthal correlations that mimic the CME signature but are distinct from backgrounds tied to elliptic flow.

Our calculations indicate that the net contribution from coherent $\rho^0$ photon production is negative, with a magnitude of $\langle \Delta\gamma^{\mathrm{coherent}\ \rho^0} \rangle \approx -0.33 \times 10^{-6}$. This corresponds to approximately $-0.2\%$ of the inclusive $\Delta\gamma$ measurement reported in semi-central Au+Au collisions at $\sqrt{s_{\rm NN}} = 200$ GeV. 
Consequently, the current experimental measurement of the possible CME signal is underestimated by $0.2\%$ in $f_{\rm CME}$.

Fortunately, this background can be effectively isolated from the signal due to its distinct kinematic signature: an extremely low transverse momentum ($p_T < 100$ MeV/$c$), determined by the nuclear size scale. We therefore recommend the application of a lower cut on the pair transverse momentum, e.g., pair $p_T > 100$ MeV/$c$, in future analyses. This strategy can significantly mitigate the coherent $\rho^{0}$ background and improve the robustness of CME measurements, such as the new high statistics of Au+Au collisions at $\sqrt{s_{\rm NN}}$ = 200 GeV at RHIC.

The key idea of the RHIC isobar program (Ru+Ru and Zr+Zr collisions) is to create systems with different magnetic field strengths by colliding nuclei with different proton numbers $Z$ but the same mass number, which is expected to result in different CME signals~\cite{STAR:2023gzg,Kharzeev:2022hqz}.
Meanwhile, the photonuclear process is also directly related to the nuclear charge $Z$, but it contributes as a background. Since the $\rho_0$ mesons produced from photoproduction introduce a negative contribution to the $\Delta\gamma$ observable, this background would lead to a larger negative contribution in Ru+Ru than in Zr+Zr collisions. This would correspond to a smaller $\Delta\gamma$ in Ru+Ru than in Zr+Zr, consistent with the experimental observations qualitatively.
More quantitatively, the photon flux approximately scales with $Z^2$, leading to a $\approx20\%$ difference in the $\rho_0$ contribution between the two isobar species. 
At $\sqrt{s_{NN}}=200$~GeV, the $\rho_0$ contribution to $\Delta\gamma$ in the isobar system would be comparable to that in Au+Au collisions. 
The resulting $\Delta\gamma$ difference from photoproduction is therefore estimated to be $\approx0.04\%$ in the isobar collisons, 
while the experimentally observed $\Delta\gamma$ difference is $\approx3\%$.

When considering the beam-energy dependence, the photon flux density approximately scales with $\ln(\gamma)$, where the Lorentz factor $\gamma = E_{\rm beam}/m_N$. 
In addition, the $\sigma(\gamma A \rightarrow VA)$ cross section for $\rho$ meson production~\cite{Klein:1999qj} shows only a weak energy dependence. 
Therefore, at $\sqrt{s_{NN}}\approx20$~GeV, the total $\rho$ photonuclear cross section is expected to be roughly a factor of $\approx2$ smaller than that at 200~GeV.
After accounting for the multiplicity dilution factor, where the $\pi$ yield roughly scaled with $\ln(E_{\rm beam})$~\cite{Chen:2024aom}, the $\langle\Delta\gamma_{\rm bkgd}^{\rm coherent\,\rho^{0}}\rangle$ is estimated to be $\approx2$ times larger than that at 200~GeV. 
Experimentally, the inclusive $\Delta\gamma$ is observed to be $\approx1.4$ times larger at 20~GeV than at 200~GeV~\cite{STAR:2014uiw,STAR:2025uxv,STAR:2025vhs}. 
This suggests that the relative contribution from $\rho_0$ photoproduction to $f_{\rm CME}$ could be of similar magnitude at the two energies, where the experimental data indicate $f_{\rm CME}\approx15.9\%$ around 20~GeV~\cite{STAR:2025vhs}.

As discussed in ref.~\cite{Deng:2014uja}, the strong electric fields generated in asymmetric Cu+Au collisions can induce nontrivial effects, 
where the in-plane electric field would drive a sizable in-plane charge dipole. 
If the electric field persists long enough, a similar mechanism could influence the photoproduction process considered here. 
Since the $\rho_0$ mesons produced from photoproduction already introduce a negative contribution to the $\Delta\gamma$ observable, the presence of the in-plane electric field in Cu+Au collisions could further enhance the magnitude of this contribution.

\section{Acknowledgments} 
J. Z thanked Dr. Fuqiang Wang for useful discussions. 
This work is supported in part by the National Key Research and Development Program of China under Contract No. 2022YFA1604900, by the National Natural Science Foundation of China under Contract No. 12275053, No. 12147101, and No. 12547102, and by the Natural Science Foundation of Shanghai under Grant No. 23JC1400200.
\end{CJK}

\bibliography{ref}

\end{document}